\documentclass[twocolumn,showpacs,preprintnumbers,amsmath,amsfonts,prl]{revtex4}
\preprint{
\hskip189pt LA-UR:04-6267}            

\usepackage{graphicx}
\usepackage{dcolumn}
\usepackage{bm}

\usepackage{color}
\definecolor{red}{rgb}{1,0,0}
\definecolor{green}{rgb}{0,1,0}
\definecolor{blue}{rgb}{0,0,1}

\begin{document}


\title{Direct observation of the formation of polar nanoregions in Pb(Mg$_{1/3}$Nb$_{2/3}$)O$_3$
using neutron pair distribution function analysis}

\author{I.-K. Jeong}
\altaffiliation {} \email{jeong@lanl.gov}
\author{T. W. Darling}
\author{J. K. Lee}
\author{Th. Proffen}
\author{R. H. Heffner}
\affiliation{Los Alamos National Laboratory, Los Alamos, NM 87545,
USA.}

\author{J. S.  Park}
\author{K. S. Hong}
\affiliation{School of Materials Science and Engineering, Seoul
National Univ., Seoul, Korea.}
\author{W. Dmowski}
\affiliation{Department of Materials Science and Engineering,
Univ. Tennessee, Knoxville, TN 37996, USA}
\author{T. Egami}
\affiliation{Department of Materials Science and Engineering and
Department of Physics and Astronomy, Univ. Tennessee, Knoxville,
TN 37996 and Oak Ridge National Laboratory, Oak Ridge, TN, 37831,
USA}


\begin{abstract}

Using neutron pair distribution function (PDF) analysis over the
temperature range from 1000~K to 15~K, we demonstrate the
existence of local polarization and the formation of medium-range,
polar nanoregions (PNRs) with local rhombohedral order in a
prototypical relaxor ferroelectric Pb(Mg$_{1/3}$Nb$_{2/3}$)O$_3$.
We estimate the volume fraction of the PNRs as a function of
temperature and show that this fraction steadily increases from 0
\% to a maximum of $\sim$ 30\% as the temperature decreases from
650~K to 15~K. Below T$\sim$200~K the PNRs start to overlap as
their volume fraction reaches the percolation threshold.
We propose that  percolating PNRs and their concomitant overlap
play a significant role in the relaxor behavior of
Pb(Mg$_{1/3}$Nb$_{2/3}$)O$_3$.

\end{abstract}

\maketitle

Relaxor ferroelectrics, such as lead magnesium niobate
Pb(Mg$_{1/3}$Nb$_{2/3}$)O$_3$ (PMN), have been widely applied due
to their high, relatively temperature-independent, permittivities.
Relaxors are characterized by a frequency dispersion in their
maximum permittivity temperature (T$_{\rm M}\sim$285~K in PMN),
with no macroscopic phase transition into a ferroelectric state at
T$_{\rm M}$~\cite{smolenskii;sptp58}. These behaviors are
fundamentally different from those of normal ferroelectrics and
are similar to spin-glasses~\cite{binder;rmp86}. It is now
believed that local polarization, resulting from structural and
chemical disorder, plays a crucial role in relaxor
behavior~\cite{cross;ferro87,samara;jpcm03}. Several models have
been proposed to explain the nature of the interactions between
these local polarizations and the mechanism for relaxor
behavior~\cite{cross;ferro87,viehland;prb91,colla;prl95,westphal;prl92,
you;prl97}. These models include
superparaelectric~\cite{cross;ferro87,viehland;prb91} and dipole
glasses~\cite{colla;prl95}, as well as random field
interactions~\cite{westphal;prl92}. The microscopic mechanism
behind  relaxor ferroelectricity is, however, not fully
understood, and is still controversial due to its high degree of
complexity.

In this Letter, we report the temperature evolution of the local
and medium-range crystal structure of PMN from 1000~K to 15~K
using neutron pair distribution function (PDF) analysis. We
present evidence for both local atomic displacements (local
polarization) and for medium-range ($\sim$ 5~{\AA} - 50~{\AA})
ordering, called polar nanoregions (PNRs). These medium-range
correlations are modeled using rhombohedral symmetry, enabling for
the first time an estimate of the temperature dependence of the
volume fraction of the PNRs. This fraction increases with
decreasing temperature, reaching the three dimensional percolation
threshold below T$\sim$200~K, where the PNRs start to overlap.  We
discuss the implications of these findings on the dielectric
properties of PMN.

Evidence for the existence of local polarization in PMN below
T$_{\rm d}\sim 620$~K, known as the Burns
temperature~\cite{samara;jpcm03}, was deduced from  optical  and
strain measurements ~\cite{burns;ssc83,cross;ferro87,zhao;apl98}.
The atomic nature of this polarization in PMN was also
studied using  extended x-ray
absorption fine structure
(EXAFS)~\cite{prouzet;jpcm93,chen;jpcs96} and atomic pair
distribution function (PDF)
analysis~\cite{egami;ferro91,rosenfeld;ferro93}.
Nb K-edge EXAFS measurements on PMN suggested a displacement of Nb
ions with respect to their oxygen octahedra by about 0.1 {\AA} from room
temperature down to 4.5~K. Pb K-edge EXAFS measurements indicated
strong static disorder on the Pb sites, but provided no detailed
information about  static Pb
displacements~\cite{prouzet;jpcm93}.
This is in contrast to neutron PDF measurements on PMN which
clearly showed that the Pb ions are off-center with respect to the
O$_{12}$ cage by as much as
0.5~{\AA}~\cite{egami;ferro91,rosenfeld;ferro93}.
In these earlier PDF measurements, however, a complete picture of
local structure in PMN was not obtained due to the limited
momentum transfer $Q$ and temperature range available. Thus, the
nature of PNRs and their volume fraction  were not determined.

Our measurements were performed on the NPDF instrument at the Los
Alamos Neutron Science Center (LANSCE).  Powder diffraction
patterns  were corrected for background, absorption and
multiple scattering, and normalized using a vanadium spectrum to
obtain the total scattering structure function $S(Q)$, using the
PDFgetN program~\cite{proffen;jac99}.
The PDF $G(r)$ is obtained from $S(Q)$ via the Fourier transform
shown in Eq. 1,
\begin{equation}
G(r)=4\pi r [\rho(r)-\rho_0]={2 \over \pi} \int_0^{Q_{max}} {\rm
Q[S(Q)-1]sinQr\,dQ}, \label{eq;pdf}
\end{equation}
where $\rho$(r) and $\rho_0$ are the atomic number and average number
densities, respectively.
Since $S(Q)$ includes both Bragg  and diffuse scattering, the
resultant PDF provides short-, medium-, and long-range structural
information~\cite{egami;bk03}. This technique has been used to
study local atomic structures~\cite{jeong;prb01} and correlated
atomic motions of atom pairs~\cite{jeong;jpca99} in many
materials.
\begin{figure}[h]\vspace{-0cm}
\includegraphics[angle=0,scale=0.9]{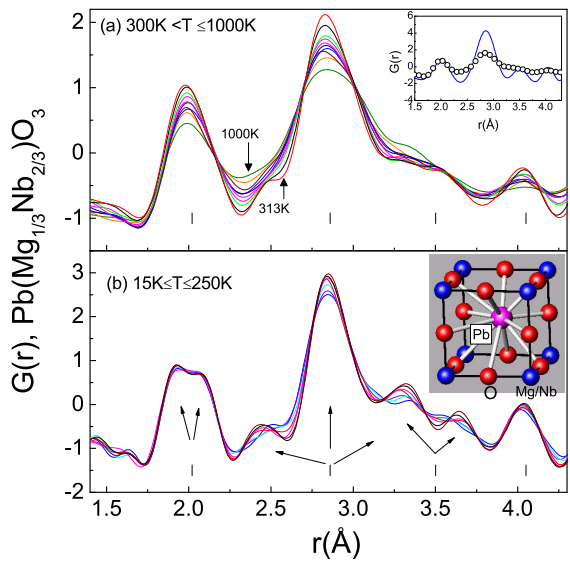}
\caption{Temperature evolution of PDF spectra of PMN within a unit
cell. (a) High temperature : T=1000~K, 850~K, 750~K, 700~K, 650~K,
600~K, 550~K, 500~K, 400~K, and 313~K, with Q$_{\rm max}$ = 25
{\AA}$^{-1}$. The inset shows a comparison between an experimental
PDF (symbol) and a model PDF (line) at 650~K. See text for
details. (b) Low temperature : T=250~K, 220~K, 190~K, 150~K,
100~K, 50~K, and 15~K, with Q$_{\rm max}$~=~30~{\AA}$^{-1}$. The
tick marks indicate (from left to right, respectively) Mg/Nb-O,
(Pb-O, O-O), Mg/Nb-Pb, and (Pb-Pb, O-O, Mg/Nb-Mg/Nb) bond lengths
in an ideal cubic perovskite structure of PMN. The inset shows the
cubic crystal structure with the Pb ion at body center.}
\label{fig;fig1}
\end{figure}
We first examine the temperature evolution of the low-$r$ region
of the PDF spectra. Fig.~1~(a) shows  spectra  from 1000~K to
313~K. The tick marks indicate the PDF peak positions for an ideal
cubic perovskite structure. The first tick mark represents
Mg/Nb-O, the second Pb-O, O-O, the third Mg/Nb-Pb, and the fourth
Pb-Pb, O-O, Mg/Nb-Mg/Nb bond lengths.
At high temperatures  note that the peaks are quite broad and the
Pb-Mg/Nb bond is not well defined due to an overlap with the Pb-O
bond. As the temperature decreases to room temperature, the Pb-O
peak gradually resolves into two peaks  at $\sim$~2.86~{\AA} and
2.5~{\AA}. This splitting indicates a displacement of Pb ions in
their O$_{12}$ cages, which breaks 12-fold degeneracy of Pb-O bond
distance (see the inset of Fig~1(b)).

In order to understand the origin of these phenomena we compared
the experimental PDF at 650~K with a model PDF at the same
temperature (shown in the inset of Fig.~1~(a)). In the model PDF
we assumed no static displacements and calculated the widths using
thermal parameters for Pb, Mg/Nb and O ions obtained from our
Rietveld refinement at 650~K.
Since the area of the Pb-O peak should be conserved, the
comparison shown in the inset to Fig.~1 indicates that the Pb-O
peak is much broader in the experiment than in the model, implying
that significant static Pb local displacement exists already at
650~K.
Thus, the gradual splitting of the Pb-O peak at lower temperatures
is mainly due to a decrease of thermal vibrations. Static Pb
displacements in PMN above the Burns temperature have also been
deduced from Raman measurements~\cite{siny;ferro89}.

Below room temperature~(Fig.~1~(b)) the experimental PDFs reveal
more interesting structure. First, the Mg/Nb-O peak around
2.02~{\AA} splits into two peaks, suggesting local Mg/Nb
displacements along the $\langle 111 \rangle$ direction. Note that
this peak shows almost no temperature dependence between 250~K and
15~K,  consistent with Nb K-edge EXAFS
measurements~\cite{prouzet;jpcm93}. In general, PDF peaks become
sharper with decreasing temperature due to  decreasing thermal
motion. Thus, this unusual temperature dependence indicates an
increasing distribution of static Mg/Nb displacements with
decreasing temperature.
Second, the Pb-O peak around 2.85~{\AA} splits into three peaks at
$\sim$  2.45~{\AA}, 2.85~{\AA}, and 3.33~{\AA}.
Third, the Pb-Mg/Nb peak around 3.5~{\AA} splits into two Pb-Mg/Nb
bonds at $\sim$~3.33~{\AA} and 3.64~{\AA} mostly due to the
displacement of Pb ions. Thus, the peak around 3.3~{\AA} has both
Pb-O and Pb-Mg/Nb components. The splitting of the Pb-O and
Pb-Mg/Nb peaks below room temperature suggests that the Pb ions
may locally shift mostly along $\langle 100\rangle$ or $\langle
111\rangle$ directions.
Interestingly,  the peak around 4.05~{\AA} (lattice constant of
PMN) is well defined; thus, the overall cubic symmetry is
maintained although the local cubic symmetry is not.
\begin{figure}[h]\vspace{-0cm}
\includegraphics[angle=0,scale=0.9]{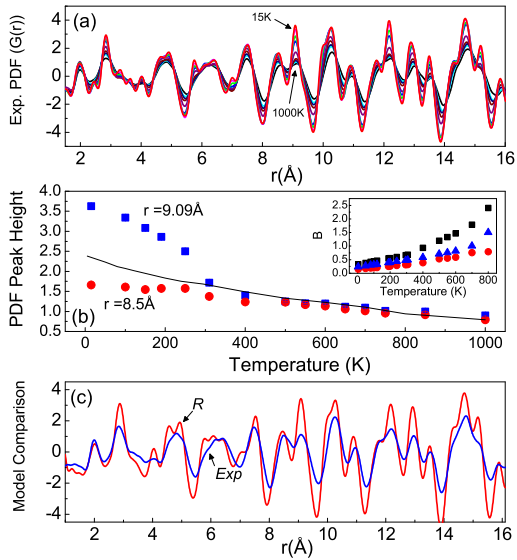}
\caption{(a) Temperature dependence of the PDF spectra of PMN at
temperatures T=1000~K, 750~K, 650~K, 500~K, 313~K, 250~K, 190~K,
100~K, 15~K. All PDFs were obtained with Q$_{\rm
max}$=25{\AA$^{-1}$}. (b) PDF peak height in the doublet between
8.0~{\AA} $<r<$9.8~{\AA} as a function of temperature. The line is
the expected average PDF peak height using the thermal parameters
in the inset. The inset shows the thermal parameters of Pb
(square), Mg/Nb (circle) and O (triangle)~\cite{bonneau;jssc91}.
(c) Comparison of rhombohedral (R-model) PDF with the experimental
PDF at 650~K. $R$ and $Exp$ represent the R-model and experimental
(paraelectric cubic) PDFs, respectively.}
\label{fig;fig2}
\end{figure}
We now address the medium-range correlations (PNRs) between these
local atomic displacements. Fig.~2~(a)  shows the PDF spectra from
1000~K to 15~K in the $r$-range up to 16~{\AA}, and
Fig.~2 (b) shows the temperature dependence of the peak
intensities in the doublet between 8.0~{\AA} $<r<$9.8~{\AA}. Here
both peaks have contributions from almost all possible ionic
pairs. The solid line is the calculated temperature dependence of
the peak height using the average Pb, Mg/Nb, and O  thermal
parameters  from Bonneau {\it et
al.}~\cite{bonneau;jssc91}, which are reproduced in the inset.
Note the strong deviation of the data below T$\sim$250~K, an indication of
PNRs, as we now show.

We modeled the PNRs using  rhombohedral symmetry (R-model: space
group R3m), wherein the Pb, Mg/Nb and oxygen octahedra are
displaced along the $\langle 111\rangle$ directions, assuming
rigid oxygen octahedra.
Similar models of rhombohedral correlations with $\langle
111\rangle$ atomic
displacements~\cite{mathan;jpcm91,takesue;prb01,blinc;prl03} and
non-collinear $\langle 100 \rangle$ displacements that average to
rhombohedral symmetry~\cite{dmowski;jpcs00} have been proposed in
Pb(Mg$_{1/3}$Nb$_{2/3}$)O$_3$ and Pb(Sc$_{1/2}$Ta$_{1/2}$)O$_3$,
respectively.
Attempts to model our data using displacements of
orthorhombic symmetry (with the atoms and oxygen octahedra
displaced along $\langle 110\rangle$ directions with the same  magnitudes as in the
rhombohedral model) were unsuccessful, with some PDF peaks being too strong and others too
weak.

In Fig.~2~(c) we compare the R-model PDF with the experimental PDF
at 650~K. For the R-model PDF calculation the thermal parameters
were taken from Bonneau {\it et al.}~\cite{bonneau;jssc91}, and
the following atomic positions were used, determined to best
simulate the differences between low temperature PDFs and the PDF
at 650~K: Pb (-0.0392, -0.0392, -0.0392), Mg/Nb (0.5062, 0.5062,
0.5062) and O (0.5308, 0.5308, 0.0308).

The PDF peaks of the R-phase are
quite distinct from those of the paraelectric cubic phase. In fact, the
relative intensities of the PDF peaks in the R-phase and in the paraelectric
cubic phase resemble the low-temperature and high-temperature
end members shown in Fig.~2 (a), respectively.
Therefore, if we assume that PNRs are dispersed as ``islands" in
the paraelectric cubic lattice~\cite{mathan;jpcm91},  growing in
size and volume fraction with decreasing temperature, we
expect that features of the  R-model PDF will become
more evident with decreasing temperature. This is exactly what we
observe in the experimental PDFs shown in Fig.~2~(a).
\begin{figure}[h]\vspace{-0cm}
\includegraphics[angle=0,scale=0.8]{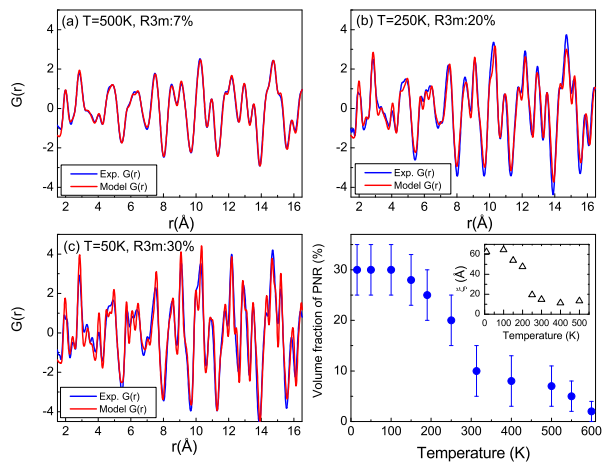}
\caption{Comparison of the ``two-phase" model PDF with the
experimental PDFs at (a) T=500~K, (b) T=250~K, (c) T=15~K.  The
model PDF is calculated from G$_{\rm m}$~=~$\alpha$(T)G$_{\rm
R}$+(1-$\alpha$(T))G$_{\rm C}$, where G$_{\rm R}$ and G$_{\rm C}$
are model PDFs for the rhombohedral and paraelectric phases.
$\alpha$(T) is the volume fraction of the rhombohedral phase. (d)
The volume fraction increases with decreasing temperature,
reaching 30\% at 15K. The inset shows the correlation length of
local polarization as a function of the temperature (taken from Xu
{\it et al.}~\cite{xu;prb04}).} \label{fig;pmn-pdf-temp}
\end{figure}

The volume fraction of the PNRs was estimated using a simple model
PDF: G$_{\rm m}$ = $\alpha$(T)G$_{\rm R}$+(1-$\alpha$(T))G$_{\rm
C}$, where G$_{\rm R}$ and G$_{\rm C}$ are rhombohedral and
paraelectric cubic model PDFs, respectively, and $\alpha$(T) is
the volume fraction of the rhombohedral phase as a function of
temperature.
For the R-model PDF calculations the atomic coordinates given
above were used for all temperatures. For the paraelectric cubic
phase, we first introduced static displacements of ions in a
$10\times10\times10$ unit cell via a reverse Monte-Carlo (RMC) fit
to the the experimental PDF at 650~K, using the DISCUS
program~\cite{proff;jac97}.
These static displacements were then fixed, and the cubic model
PDF was calculated at various temperatures using the thermal
parameters  from Bonneau {\it et al.}~\cite{bonneau;jssc91}. (The
same thermal parameters are used for both cubic and rhombohedral
phases.) After  calculating both paraelectric cubic and
rhombohedral PDFs at a given temperature, $\alpha$(T) was adjusted
to obtain the best match to the corresponding experimental PDF.
Figures~3~(a-c) show these comparisons at 500~K, 250~K, and 50~K
in the $r$-range up to $r$=16~{\AA}. Considering the simplicity of
the model PDF calculations, the overall agreement at various
temperatures is very good.

In Fig.~3(d) we show the temperature dependence of $\alpha$(T),
along with the correlation length of local polarization determined
by Xu {\it et al.} using neutron elastic diffuse
scattering~\cite{xu;prb04}, shown in the inset.  The volume
fraction steadily increases with decreasing temperature, reaching
$\sim$ 30\% at 15~K. The correlation length $\xi$ of local
polarization increases from $\xi\sim$~15~{\AA} to
$\xi\sim$~60~{\AA}.
Note that around T$\sim$200~K, where the correlation length
drastically increases, $\alpha$(T) reaches the percolation
threshold P$_c$$\sim$28\% for spherical objects in three
dimensions~\cite{garboczi;pre95}. This implies that the PNRs start
to overlap with each other below T$\sim$200~K. This correlates
well with the abrupt increase in the correlation length of local
polarization to $\xi\sim$~50 - 60~{\AA}, determined independently.

Neutron scattering measurements by Wakimoto {\it et al.}
~\cite{wakimoto;prb02} suggest that locally, each PNR maintains a
stable spontaneous polarization below T$\sim$220~K where the
correlation length is about $\sim$ 50-60 {\AA}.
Further growth of these large PNRs (beyond $\sim$~60~{\AA}) may be
hindered by electric and elastic energy barriers, which are
proportional to the size of the PNR. In an applied electric field,
however, the polarizations in the separate PNRs are aligned, and
the separate PNRs form a macro polar domain. Once the macro polar
domain is formed, the dipoles in the domain do not randomly orient
even after the electric field is turned off.

This picture helps to explain some basic behavior found in PMN,
such as hysteresis in the field-induced
polarization~\cite{bokov;spss61}, the field-induced rhombohedral
phase transition below 220 K~\cite{calvarin;ferro95} and the
anomaly in the dielectric permittivity near 212 K upon field
heating after zero-field cooling~\cite{westphal;prl92}.
Our measurements, therefore,  provide direct  local structural
evidence for the importance of percolating PNRs in explaining many
features of the relaxor behavior found in PMN below the Burns
temperature.

\acknowledgments

Work at Los Alamos was carried out under the auspices of the US
DOE/Office of Science. This work has benefited from the use of
NPDF at the Lujan Center at Los Alamos Neutron Science Center,
funded by DOE Office of Basic Energy Sciences and Los Alamos
National Laboratory funded by Department of Energy under contract
W-7405-ENG-36. The upgrade of NPDF has been funded by NSF through
grant DMR 00-76488.


\end{document}